\documentclass[universe,review,accept,pdftex,moreauthors]{Definitions/mdpi} 

\firstpage{1} 
\pubvolume{1}
\issuenum{1}
\articlenumber{0}
\pubyear{2024}
\copyrightyear{2024}
\externaleditor{Academic Editor: }
\datereceived{29 February 2024} 
\daterevised{21 April 2024} 
\dateaccepted{ } 
\datepublished{ } 
\hreflink{https://doi.org/} 

\Title{SUPERNOVA REMNANTS IN GAMMA RAYS}

\TitleCitation{Supernova Remnants in Gamma Rays}

\renewcommand{\hl}{}     

\Author{Andrea Giuliani 
 $^{1,}$*\orcidA{
https://orcid.org/0000-0002-4315-1699} and Martina Cardillo 
 $^{2}$\orcidB{https://orcid.org/0000-0001-8877-3996}}

\AuthorNames{Andrea Giuliani, Martina Cardillo}

\AuthorCitation{\hl{Giuliani, A.;} 
 Cardillo, M.}

\address{%
$^{1}$ \quad INAF---Istituto di Astrofisica Spaziale e Fisica Cosmica,
  Via Alfonso Corti 12, 20133 Milano, Italy; andreaa.giuliani@inaf.it 
\\
$^{2}$ \quad INAF---Istituto di Astrofisica e Planetologia Spaziali,
  Via del Fosso del Cavaliere 100, 00133 Roma, Italy; martina.cardillo@inaf.it
  }

\corres{Correspondence: andrea.giuliani@inaf.it}

\abstract{In
 the 1960s, the remnants of supernova explosions (SNRs) were indicated as a possible source of galactic cosmic rays through the Diffusive Shock Acceleration (DSA) mechanism. Since then, the observation of gamma-ray emission from relativistic ions in these objects has been one of the main goals of high-energy astrophysics. A few dozen SNRs have been detected at GeV and TeV photon energies in the last two decades. However, these observations have shown a complex phenomenology that is not easy to reduce to the standard paradigm based on DSA acceleration. Although the understanding of these objects has greatly increased, and their nature as efficient electron and proton accelerators has been observed, it remains to be clarified whether these objects are the main contributors to galactic cosmic rays. Here, we review the observations of $\gamma$-ray emission from SNRs and the perspectives for the future. }


\def \gray {$\gamma$-ray}
\def \grays {$\gamma$-rays}

\def \apj {Astrophys. J.}
\def \apjl {Astrophys. J. Lett.}
\def \apjs {Astrophys. J. Suppl. Ser.}
\def \aap {Astron. Astrophys.}

\def \jcap {J. Cosmol. Astropart. Phys.}
\def \mnras {Mon. Not. R. Astron. Soc.}

\def \pasj {Publ. Astron. Soc. Jpn.}

\def \prd {Phys. Rev.}

\def \nat {Nature}

\def \prl {Phys. Rev. Lett.}
\def \aapr {Astron. Astrophys. Rev.}

\begin{document}

\section{Supernova~Remnants}
\label{sec:general_SNRs}
\unskip

\subsection{Supernova Remnant~Evolution}
\label{sec:SNRs}

Supernova explosions can result from two main physical processes related to the life cycle of a star. 
The first one happens when a star with a mass greater than 8 ${M}_\odot$ reaches the end of its life, and the radiation produced within the nucleus is no longer able to counteract the weight of the star's envelope. 
The collapse of the star triggers the material to produce a supernova explosion.
Alternatively, a~white dwarf in a binary system can exceed the Chandrashekar equilibrium limit (approximately 1.4 ${M}_\odot$) due to accretion from its companions and explode in a type Ia supernova.
In both cases, the~supernova explosion expels a significant amount of matter, ranging from 1 to several solar masses, into~the interstellar medium at a speed of $10^4$ km s$^{-1}$. This corresponds to a kinetic energy of approximately $10^{51}$ erg.
The main differences between the two scenarios concern the chemical composition of the ejected materials and the interstellar medium (ISM) into which the supernova remnant (SNR) expands.

Over the subsequent 100,000 years, the~ejecta will expand, interact with the surrounding medium, and~eventually dissolve into it.
The evolutionary phases of an SNR can be defined by the type of interaction with the surrounding medium and the ratio between the ejecta mass ($M_{\rm ej}$) and the mass of the swept-up material ($M_{\rm sw}$).

During the  {free expansion} phase, lasting a few hundred years as long as $M_{\rm ej} > M_{\rm sw}$, the~star's ejecta expands freely into the surrounding medium with an expansion law linear in time, $R \sim t$.
In this phase, the~material's speed exceeds the speed of sound in the surrounding medium, thus creating a shock.
The temperature decreases as the gas expands adiabatically: $T \sim -3(\gamma-1)$, where $\gamma$ is the specific heat ratio.
 
In the {\hl{Sedov--Taylor phase}
}, lasting between 20,000 and 40,000 years as long as $M_{ej} \leq M_{sw}$, the~deceleration of the shell becomes significant, and~a reverse shock begins to propagate toward the interior of the SNR, heating the gas, which thus becomes visible in the soft X-ray band.
The evolution of the remnant is well described by the adiabatic blast-wave solution of Taylor and Sedov~\cite{Sedov59}, and the radius increases with time as $R \propto t^{2/5}$. 

The forward and reverse shocks formed during the  {free expansion} and  {Sedov--Taylor} phases can accelerate particles (see Section~\ref{accel}).
The observation of non-thermal radio and \gray\ emissions from SNRs with ages up to a few tens of thousands of years confirms this~aspect. 

\begin{figure}[H]
    \includegraphics[width=0.7\textwidth]{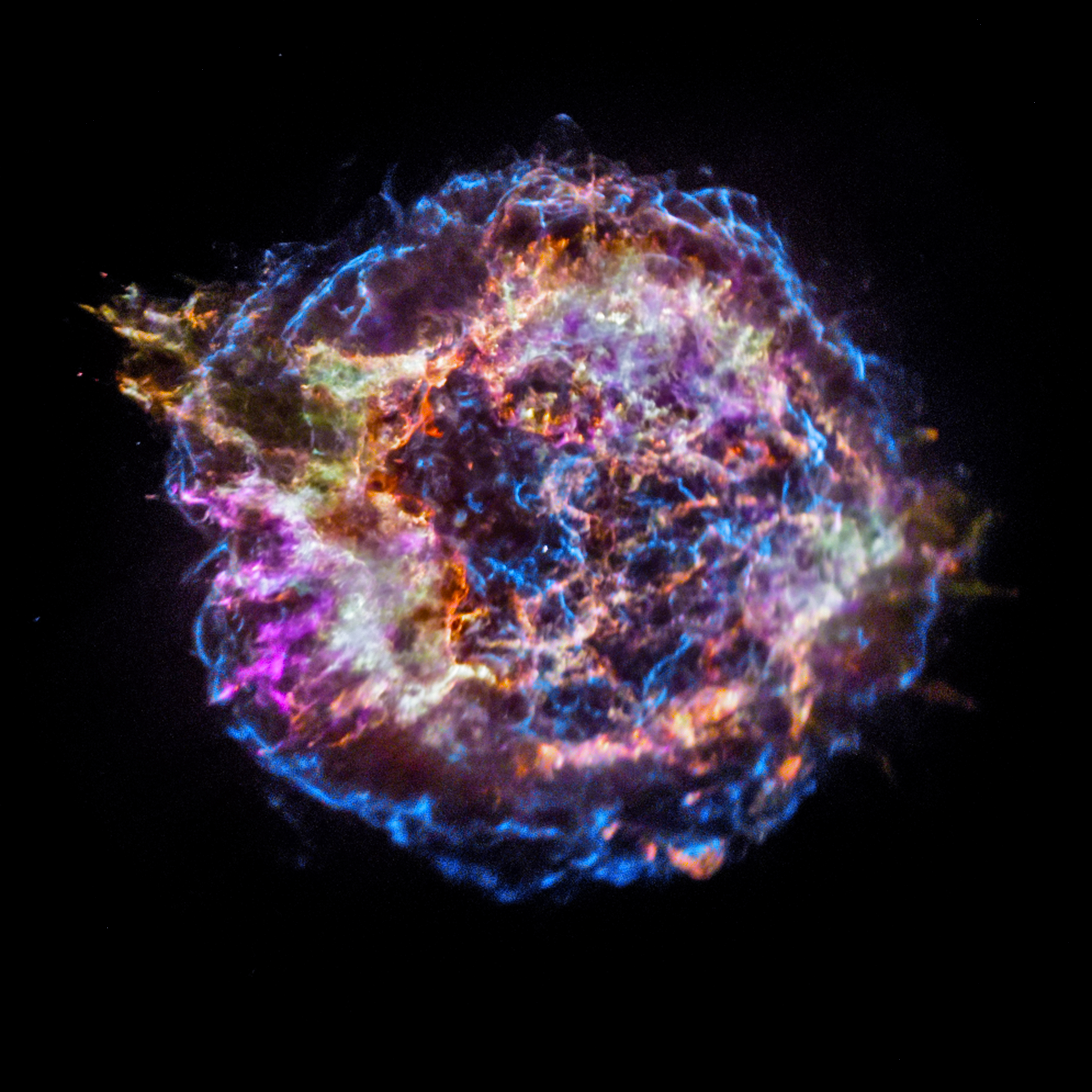}
    \caption{\hl{Cas} 
 A image in X-ray band as observed by the Chandra X-Ray Observatory telescope \citep{lee14_casAwithChandra}.}
    \label{fig:CasA_X}
\end{figure}

In the  {radiative phase} ($M_{\rm ej} \ll M_{\rm sw}$), the shell expands at a sub-thermal speed.  
After this phase, the~remnant of the supernova dissolves and becomes part of the interstellar~medium.

The duration of each phase of an SNR is determined by the interaction between the ejecta and the ISM, which, in turn, depends on the type of circumstellar medium in which the SNR is situated.
On the other hand, supernova explosions play a crucial role in shaping the evolution of the ISM and galaxies by injecting a significant amount of energy and momentum and then influencing the process of star formation~\cite{Koo20_SNRmomentum}.
Figure \ref{fig:CasA_X} shows the young SNR Cas A as seen in the X-ray. 


\subsection{Supernova Remnants as Cosmic~Accelerators}
\label{accel}

In the initial two stages of an SNR's evolution, a~shock forms between the ejecta (including swept material) and the local interstellar medium. Additionally, a~second shock may develop and propagate toward the~center.

Over the past decades, various works have shown how shocks can accelerate particles, including cosmic rays (CRs), and~there is a consensus that the dominant process is Diffusive Shock Acceleration (DSA) \cite{Bell78, Blandford78}. This process is based on Fermi I-order acceleration \citep{Fermi49}.
The acceleration index provided by DSA, $\gamma= \frac{3R}{R-1}$, is strictly correlated with the compression ratio of the shock, $R=\frac{u_d}{u_u}$, where $u_u$ and $u_d$ are the upstream and downstream velocities, respectively, related to the shock Mach number. For~strong shocks and in the test-particle limit, i.e.,~when the particle energy is negligible compared to that of the shock, DSA predicts a particle energy
distribution with an energy spectral index close to 2. 
However, the backreaction of CRs on the shock and then also on the surrounding ambient can modify the acceleration spectrum, and if~the feedback effects on the shock are not negligible, the~spectral index of accelerated particles may become steeper than 2
(for a review of the DSA theory, see  \citep[][]{Blasi13} and references therein).

\begin{figure}[H]
    \includegraphics[width=1.\textwidth]{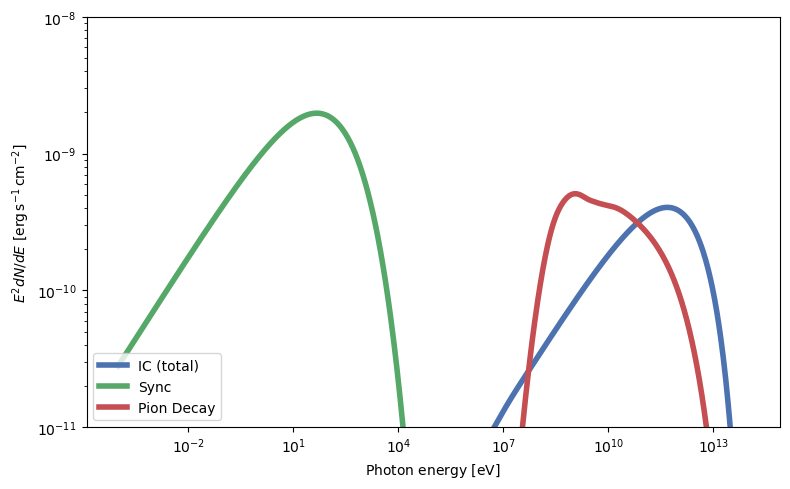}
    \caption{Example of SED given by synchrotron, IC and pion decay~process.}
    \label{fig:sed}
\end{figure}

If SNRs accelerate cosmic rays (CRs), they should be observed as sources of non-thermal emission. 
This emission results from the interaction of
electrons and hadrons with the surrounding medium.  
The interaction between electrons and magnetic fields at the site produces synchrotron radiation.
If the electron spectral distribution can be described as a power law with index $\alpha$ and maximum energy $E_{\rm max}$ (i.e., $f(E)  \propto  E^{-\alpha} exp(E/E_{\rm max})$), the~emerging emission has a spectrum with slope $\gamma = - ( \alpha +1) / 2 $, typically resulting in a hard spectrum. 
In 
the spectral energy distribution 
(SED, see Figure~\ref{fig:sed}), the~peak energy of synchrotron radiation 
can  reach the X-ray band for an $E_{\rm max}$ of few TeV.
The same electron population interacting with the radiation fields produces emission through the Inverse Compton process.  
The emission has a spectral shape similar to that of the synchrotron but shifted toward higher energies,
typically in the \gray\ band.
If a dense medium is present, electrons can also radiate efficiently through Bremsstrahlung, which concurs with the \gray\ emission of the~objects. 

If hadrons (protons or heavier nuclei) are also accelerated, they scatter inelastically against the nuclei of the medium in the so-called “pp interaction”, producing neutral pions, which \hl{rapidly} 
( $8.5  \times 10^{-17}$ s) 
decay into two \gray\ photons.   
In addition to protons, secondary electrons (produced by pp interactions) can produce \gray\ emission through IC and Bremsstrahlung (see~\cite{Huang20_secondaryElectrons} and references therein). 

SNRs have often been suggested as possible sources of CRs for~two main reasons: their observed non-thermal emission (from radio to \gray) and~a substantial match between the average power of supernova explosions in the Galaxy and the power needed to keep the galactic CR population stable.  
The energy density of CRs in the Galaxy disk is on the order of $w_{\rm CR} \simeq 1$ eV/cm$^3$.
In the standard leaky-box approximation, we assume the disk of the Galaxy as a disk with a radius of 15 kpc and a height of 1 kpc, obtaining  a total CR energy 
$ E_{CR} =  w_{\rm CR} \times V_{\rm disk} \simeq  3.34 \times 10^{55}$ erg. 
Isotope ratios in CRs can be used to estimate the typical permanence time of CRs in the plane of the Galaxy, which is approximately $t_ {\rm disk} \approx 10^7$ years.
Therefore, assuming that the galactic CR distribution has reached a steady-state condition, the~power needed to maintain the observed CR density is
$
 P_{CR} = \frac{E_{CR}}{t_{disk}} \simeq 10^{41} \; \rm{erg/s}.
$
Whichever process accelerates CRs in the Galaxy, it must therefore have a power of at least $P_{\rm CR}$.
A process with similar power is given by supernova explosions. Assuming that their average rate in our Galaxy is on the order of 1/30 years, we find that
$
 P_{SN} \simeq 1 \simeq 10^{42} \;  \rm{erg/s}
$.
The first to notice this similarity were \hl{Ginzburg and Syrovatskii} \cite{Ginzburg64}, 
 who concluded that
``SuperNovae alone could maintain the CR population provided that about 10\% of their kinetic energy is somehow converted into CRs''.
This was the first evidence pointing to supernovae (or, equivalently, SNRs) as prime candidates for CR~sources.

The observation of non-thermal emission from SNRs strongly supports the hypothesis that they could be sources of galactic CRs. Indeed, the synchrotron spectra of these objects have been observed in the radio band since the 1960s, showing the presence of magnetic field amplification \citep{Koyama95, Uchiyama07, Vink12}, and the modification of Balmer lines in the optical band due to CR presence was also observed \citep{Morlino12}. 

The majority of cosmic rays are hadrons. Therefore, evidence supporting the idea that supernova remnants can sustain the galactic cosmic-ray population must come from observations of these objects in the gamma-ray band. This is the only band where a clear indication of their presence can be found.
Moreover, the~cosmic-ray spectrum suggests that some galactic sources can accelerate particles to at least 1 PeV. 
This should result in a \gray\ spectrum up to and beyond 100 TeV, with no~cutoffs.

For this reason, SNRs have always been a primary focus of \gray\ observations. This includes both the GeV band (approximately 0.1--100 GeV), which is covered by \gray\ satellites, and~the TeV band (approximately 0.1--100 TeV), which is covered by Cherenkov telescopes.
On the one hand, we now know that even the first \gray\ instruments (SAS-2, COS-B in the 1970s, EGRET on board CGRO in the 1990s) were able to detect these objects. On~the other hand, the~limited angular resolution of these instruments (and the crowding of the galactic fields) did not allow the \gray\ emission to be associated with~SNRs. 

The first certain associations were made with Cherenkov instruments; HEGRA observed a source associated with SNR W28, while H.E.S.S. was able to resolve the shell morphology of SNR RX J1713.7-3946 \citep{Abdalla18_HESS_1713}.

In recent years, many other classes of sources have been shown to accelerate CRs (\citep{Cardillo23} and reference therein), and~consequently, it is crucial to understand whether the contribution of SNRs is dominant or~not.

In this new context, another fundamental channel is neutrino detection. This is an unquestionable hint of CR acceleration since neutrinos can be produced only by the decay of charged pions produced by p-p and p-$\gamma$ interactions \citep{Anchordoqui14}. 
Looking for neutrino detection in correspondence with PeVatron candidates can confirm the nature of \gray{}-emitting sources \citep{Celli20}. 


\section{SNR in Gamma-Ray~Band}
\label{sec:SNR_gamma}

Although an exhaustive classification of SNRs in \grays\ is not possible, certain groups of objects have been recognized over time based on their spectrum morphology and multiwavelength behavior \citep{Funk17} 
\hl{ (a list of firmly identified SNRs in gamma-ray can be found in Table ) } \ref{tab1}. 
In young (few thousand years)   {shell-like} SNRs, the~\gray\ emission comes from the shell, and often, there is a good morphological correlation between X-ray and \gray\ morphology. 
Some objects of this class are  RX J1713.7-3946, RX J0852.0-4622, RCW 86, and~SN 1006.
They have similar \gray\ spectra, composed of a hard component (index < 2) peaking around a few TeV, followed by a rapid spectrum decrease. 
They also show similar \gray\ luminosities~\citep{Acero15}.
%
%
It has been proposed that this class of SNRs is leptonic; in this scenario, the~SED can be modeled quite easily with a single electron population emitting in X-rays through a synchrotron and in \grays\ through IC \citep{Acero15}. 
However, other authors proposed hadronic models for at least some of these objects.
For example,  the~spectrum of RXJ 1713.7-3946 can be explained in terms of hadronic emission, taking into account the clumpiness of the surrounding medium \citep{fukui13_1713isHadronic}. 

Another class of sources includes the SNRs interacting with molecular clouds (MCs). 
A list of SNRs in this class can be found in \citep{Jiang10_iSNRs}. 
In these objects, the~\grays\ originate from a region of dense gas that is close to, or~in contact with, the~SNR. 
In these systems, typically, the target for the accelerating particles is the gas contained in large  MCs (more than $10^3$ $M_{\odot}$).
Some famous examples are W 44, W 28, IC 443, and~W 51C.
The \gray\ spectra of these objects have a soft index (>2.5) and are more easily observed in the GeV than in the TeV band.
The ages of these SNRs reach several tens of thousands of years, and~their \gray\ luminosities are $\geq$$10^{35}$ erg/s.
These spectra can be interpreted as \gray\ emissions from particles that diffuse in a partially ionized interstellar medium
~\cite{Zirakashvili18_iSNRspectra}.
The interaction between the SNR shock and the MC can be observed through enhanced CO(2-1)/CO(1-0) ratios, OH maser emission at 1720 MHz, and~SiO emission \citep{Slane16}. 
Additionally, this interaction can be traced using the neutral iron line induced by MeV protons in dense gas~\cite{Nobukawa18_ironLine}.

Notably, the~very young SNRs Tycho and Cas A exhibit properties that are not easily classified within these two categories, with~a spectral index that is intermediate between the~two.



\begin{table}[H]
\caption{\hl{List} 
 of SNRs firmly identified with a gamma-ray source. The Columns GeV, TeV, and PeV indicate detection, respectively, in the bands 0.1--100 GeV, 0.1--100 TeV, and $>$0.1 PeV.} \label{tab1}
\newcolumntype{C}{>{\centering\arraybackslash}X}
\begin{tabularx}{\textwidth}{CCCCC}
\toprule
\textbf{Name} & \textbf{Common Name} & \textbf{GeV} & \textbf{TeV} & \textbf{PeV} \\
\midrule
G004.5+06.8 & Kepler & \hl{Yes} 
 \citep{Acero22_Kepler} & Yes \citep{HESS2024_Kepler} & - \\
G006.4-00.1 & W28 & Yes \citep{giuliani10_w28} & Yes \citep{aharonian08_W28} & - \\
G008.7-00.1 & W30 & Yes \citep{Ajello12_W30} & - & - \\
G020.0-00.2 & - & Yes \citep{Fermi_16_Icat}& - & - \\
G023.3-00.3 & W41 & Yes \citep{HESS15_W41} & Yes \citep{HESS15_W41}& - \\
G024.7+00.6 & - & Yes \citep{Fermi_16_Icat} & Yes \citep{MAGIC19_G24} & - \\
G034.7-00.4 & W44 & Yes \citep{Giuliani11_W44} & - & - \\
G043.3-00.2 & W49B & Yes \citep{Abdo10_W49b} & Yes \citep{Brun11_W49b} & - \\
G045.7-00.4 & - & Yes \citep{Fermi_16_Icat} & - & - \\
G049.2-00.7 & W51C & Yes \citep{Abdo09_W51c}& Yes \citep{Aleksic12_W51c} & Yes \citep{Cao2024_LHAASOcat} \\
G074.0-08.5 & CygnusLoop & Yes \citep{Katagiri11_Cygnus} & - & - \\
G078.2+02.1 & $\gamma$-Cygni & Yes \citep{Lande12_GCygni} & Yes \citep{Aliu13_GCygni} & Yes \citep{Cao2024_LHAASOcat} \\
G089.0+04.7 & HB21 & Yes \citep{Pivato13_HB21} & - & - \\
G106.3+02.7 & - & Yes \citep{Xin19_G106} & Yes \citep{Acciari09_G106} & Yes \citep{Cao21}\\
G109.1-01.0 & - & Yes \citep{Castro12_G109} & - & - \\
G111.7-02.1 & CasA & Yes \citep{abdo10_casAfermi} & Yes \citep{Aharonian01_casA} & - \\
G120.1+01.4 & Tycho & Yes \citep{Giordano12_Tycho} & Yes \citep{Acciari11_Tycho} & - \\
G132.7+01.3 & HB3 & Yes \citep{Katagiri16_HB3} & - & - \\
G160.9+02.6 & HB9 & Yes \citep{Araya14_HB9} & - & - \\
G180.0-01.7 & S147 & Yes \citep{Katsuta12_S147}& - & - \\
G189.1+03.0 & IC443 & Yes \citep{Tavani10_IC443} & Yes \citep{Acciari09_IC443} & - \\
G205.5+00.5 & Monoceros Loop & Yes \citep{Liu23_G205} & - & - \\
G260.4-03.4 & PuppisA & Yes \citep{Hewitt12_PuppisA} & - & - \\
G266.2-01.2 & Vela Jr & Yes \citep{Tanaka11_Vela} & Yes \citep{Aharonian07_Vela} & - \\
G291.0-00.1 & - & Yes \citep{Fermi_16_Icat} & - & - \\
G292.0+01.8 & - & Yes \citep{Parkinson16_G292_G298} & - & - \\
G296.5+10.0 & - & Yes \citep{Araya13_G296} & - & - \\
G298.6-00.0 & - & Yes \citep{Parkinson16_G292_G298} & - & - \\
G315.4-02.3 & RCW86 & Yes \citep{Yuan14_RCW86} & Yes \citep{Aharonian09_RCW86} & - \\
G321.9-00.3 & - & Yes \citep{Nolan12_FermiIIcat}& - & - \\
G323.7-01.0 & HESSJ1534-571 & - & Yes \citep{HESS18_J1534} & - \\
G326.3-01.8 & - & Yes \citep{Devin18_G326} & - & - \\
G327.6+14.6 & SN1006 & Yes \citep{Condon17_SN1006_J1731} & Yes \citep{Acero10_SN1006} & - \\
G347.3-00.5 & RXJ1713.7-3946 & Yes \citep{Abdo11_J1713} & Yes \citep{Aharonian04_J1713} & - \\
G348.5+00.1 & CTB37A & Yes \citep{Brandt13_CTB37a} & Yes \citep{Aharonian08_CTB37a} & - \\
G348.7+00.3 & CTB37B &  Yes \citep{Xin16_CTB37b}  & Yes \citep{Aharonian08_CTB37b} & - \\
G349.7+00.2 & - & Yes \citep{Castro10_G349} & Yes \citep{HESS15_G349} & - \\
G353.6-00.7 & HESSJ1731-347 & Yes \citep{Yang14_J1731} & Yes \citep{HESS11_J1731}& - \\
G355.4+00.7 & - & Yes \citep{Fermi_16_Icat} & - & - \\
G357.7-00.1 & MSH 17-39 & Yes \citep{Castro13_G357} & - & - \\
\bottomrule
\end{tabularx}
\end{table}


\subsection{Cas~A}
\label{sec:CasA}
 
Cas A is one of the youngest known SNRs in the Galaxy, with~an estimated age of about 350 years. 
It is the remnant of a Type IIb supernova \citep{Krause08_CasA} that likely passed through a red supergiant phase \citep{Chevalier03_CasA}. 
The remnant is located at a distance of 3.4 $\pm$ 0.4 kpc \citep{Reed95_CasA}.

\begin{figure}[H]
    \includegraphics[width=0.8\textwidth]{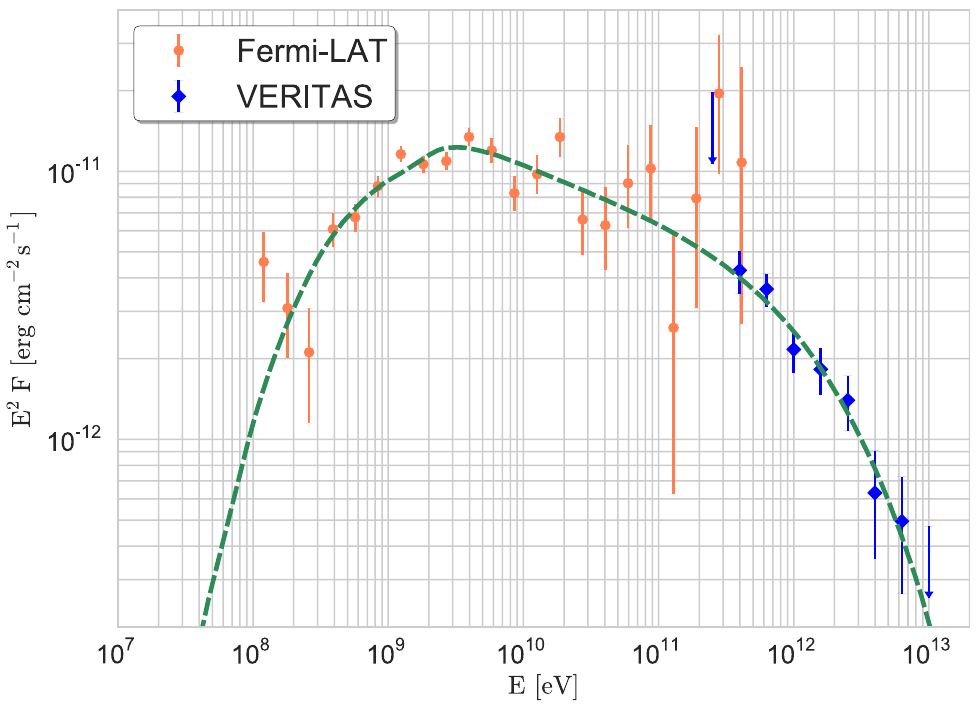}
    \caption{\hl{The Cas} 
 A spectrum, as~seen by Fermi and Veritas (Figure from \citep{abeysekara20_casAhadronic}). The~dashed line represents a hadronic model for this~source.}
    \label{fig:casA}
\end{figure}

In the \gray\ band, this SNR was extensively observed by Fermi-LAT \citep{abdo10_casAfermi}, MAGIC \citep{Albert07_CasA}, and~Veritas \citep{Acciari10_CasA}, covering an energy range from 100 MeV to about 10 TeV (see Figure~\ref{fig:casA}).
A single-zone hadronic component can accurately model the GeV-TeV spectrum (see Figure~\ref{fig:casA}).
The population of protons is described by a power law with an index of 2.17 and a cutoff energy of 17 TeV~\cite{abeysekara20_casAhadronic}. 
However, the~contribution of a leptonic component cannot be ruled~out. 
Given the observational evidence for a reverse shock in Cas A \citep{Gotthelf01_CasA, Morse04_CasA}, a~two-zone model with both forward and reverse shocks has been adopted for multiwavelength emission in many works. Moreover, multiwavelength observations indicate an asymmetric profile that could be modeled as a jet-like feature superposed onto an expanding spherical shell.
The jet-like structure can produce a \gray\ emission detectable at 100 TeV by LHAASO, CTA, and ASTRI Mini-Array \citep{zhan22_CasAModel}.

\subsection{Tycho~SNR}
\label{sec:Tycho}

Tycho's SNR is the remnant of the Type Ia SN 1572 registered by Tycho Brahe, and it has a distance of about 3 kpc. It is one of the youngest known SNRs, with an age of 450 yrs, and it is still in its ejecta-dominated phase, with~a shock velocity of about 4000 km s$^{-1}$. 

X-ray observations by the Chandra Observatory have revealed synchrotron X-ray filaments at the shock location \citep{Hwang02_Tycho} with a very high variability \citep{Matsuda22_Tycho}, a~clear indication of the presence of an amplified magnetic field, one of the main conditions to reach~VHEs.

In the $\gamma$-ray observations, there is also no clear evidence of the presence of a cutoff because the last VERITAS results have large error bars that cannot confirm it \citep{Park15_Tycho,Archambault17_Tycho}. Its power-law spectrum, with~an index $\propto E^{-2.3}$, in agreement with the theoretical expectation \citep{Morlino12_Tycho}, points toward a hadronic origin of this~emission.

\begin{figure}[H]
    \includegraphics[width=0.7\textwidth]{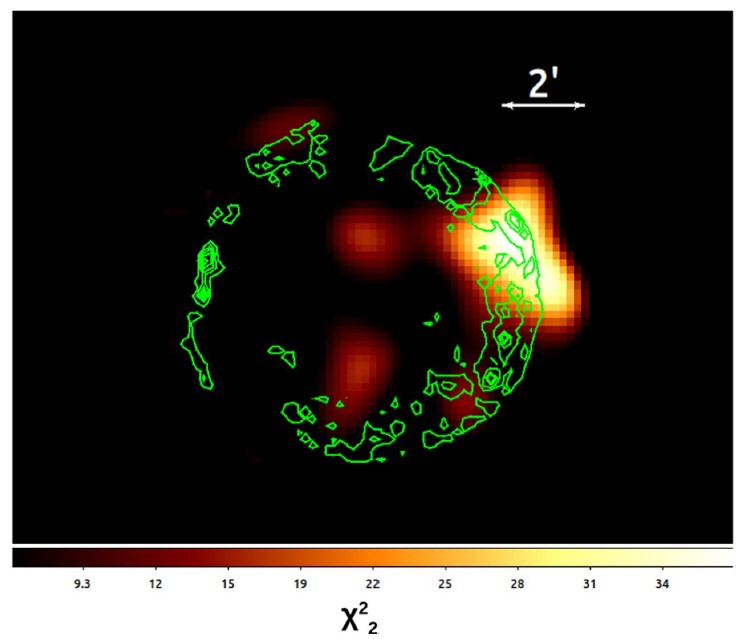}
    \caption{{The polarization} 
 map of the Tycho SNR. The color scale gives the $\chi^{2}$ values for the polarization signal in the 3–6 keV energy band smoothed with a Gaussian kernel. Superimposed in green are the Chandra 4–6 keV contours (figure from \citep{Ferrazzoli23_Tycho}).}
    \label{fig:Tycho}
\end{figure}

Very recent results obtained in X-ray polarization by IXPE \citep{Ferrazzoli23_Tycho} show a dominant radial magnetic field, in~agreement with the radio polarization one and~the one found in CasA \citep{Vink22_CasA} (see Figure~\ref{fig:Tycho}). The~polarization degree is about $12\%$, indicating the presence of such an order in the magnetic field that, however, cannot exclude the presence of~turbulence. 
\vspace{-6pt}
These features make the Tycho SNR one of the main SNR candidate PeVatrons in the CR context and one of the best studied. However, LHAASO does not include this source in its list of candidate PeVatrons, and the most recent theoretical discussions \citep{Bell13, Cardillo15} showed that the SNR could reach PeV energies only in the first 100 yrs of their~evolution. 

Only a future generation of IACTS with an effective area and a sensitivity better than those currently available will be able to better constrain the spectrum at VHEs and hence confirm or disprove the PeVatron nature of~Tycho.


\subsection{G 106.3+02.7}
\label{sec:G106}

Among the first high-significance PeVatrons published by the LHAASO collaboration~\citep{Cao21, Cardillo23}, there was only one source associated with an SNR: \hl{LHAASO J2226+6057,} 
 correlated with VER J2227+608/HAWC J2227+610 \citep{TeVCat08}. Its~estimated age is about 10,000~years at a distance of 800~pc.

\textls[-15]{Its VHE/UHE emission could be explained by two kinds of sources: the SNR G106.3+2.7} with the associated MC in~the “tail” of the TeV emission, and~the Boomerang PWN, associated with the PSR J2229+6141, collocated in the “head”. The~low resolution of the UHE detection by HAWC \citep{Albert20_G106}, Tibet ASg \citep{Amenomori21_Cygnus}, and~finally, LHAASO \citep{Cao21} does not allow us to say whether the \gray\ emission is from the head or the tail~region. 

\begin{figure}[H]
    \includegraphics[width=0.8\textwidth]{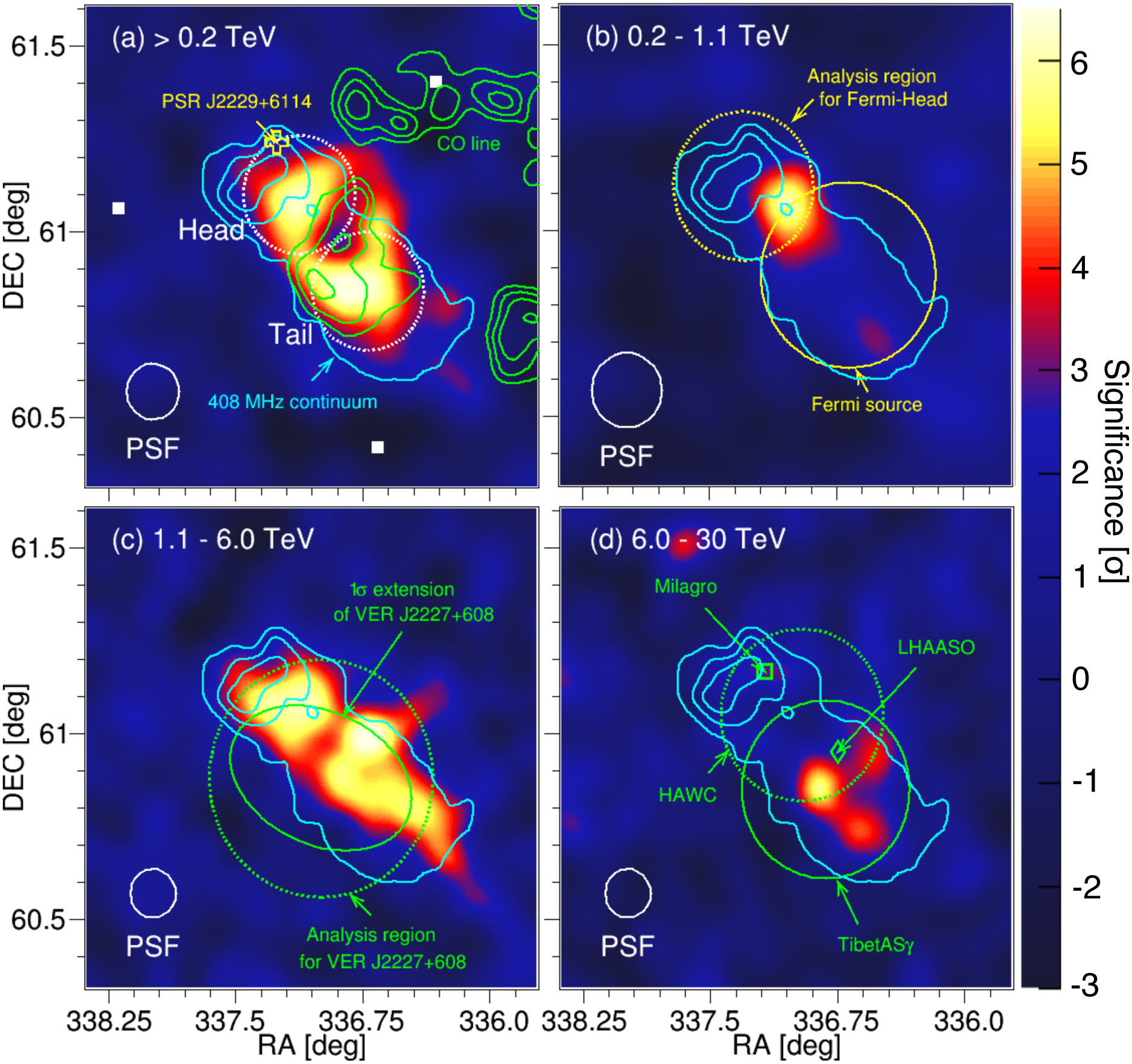}
    \caption{\hl{Energy-dependent} 
 pre-trial significance maps of the SNR G106.3+2.7 observed with the MAGIC telescopes at different energy ranges (from \citep{MAGIC23_G106}).}
    \label{fig:G106}
\end{figure}

After the second LHAASO publication \citep{Cao2024_LHAASOcat}, in which more than one SNR is a PeVatron candidate, the~chance that the VHE/UHE emission comes from the SNR G106.3+2.7 is taken into account more rigorously. This remnant was discovered by the Northern Galactic plane survey in the radio band by the DRAO \citep{Joncas90} with a comet-shaped morphology.
A recent Fermi-LAT GeV analysis showed that only the tail seems to emit at the highest energies detectable by the LAT instrument (10–500 GeV), explaining the whole emission with a hadronic model from the SNR/MC interaction \citep{Fang22_G106}. 
A hadronic origin was also declared the most likely by the MAGIC collaboration, which resolved, for~the first time, the~VHE/UHE emission (see Figure~\ref{fig:G106}), detecting $E > 10$ TeV only from the tail region \citep{MAGIC21_G106, MAGIC23_G106}. 
Several studies and analyses are ongoing to disentangle the contribution from the two possible sources (for a review, see \citep{Cardillo23}). However, the~real origin of the hadronic emission will only be understood with an in-depth analysis of the microphysics of the~region. 

In the future, the~very good angular resolution of ASTRI Mini-Array \citep{Cardillo23P, Tutone23P_G106} and CTA \citep{Verna22_G106_CTAN} could resolve the VHE emission location. 
Unfortunately, current estimations of the neutrino flux expected in the case of hadronic emission cannot be detected at the IceCube sensitivity~\cite{Sarmah23_neutrini}.


\subsection{RX~J1713.7-3946}
\label{sec:RXJ1713}

The “standard candle”  in the debate about the hadronic or leptonic origin of the  \gray\  emission from an SNR is the source RX J1713.7-3946 (G 347.3-0.5), because~its GeV-TeV emission \citep{Abdo11_J1713, Abdalla18_HESS_1713} can be reproduced with both types of models. 
Its age is about 1625~yrs with a distance of $d \simeq 1 $ kpc \citep{Fukui_1713_2003, Moriguchi_1713_2005} and an extension of $R_{\rm s}\simeq 0.6$ deg \citep{Abdalla18_HESS_1713}.

\begin{figure}[H]
    \centering
    \includegraphics[width=0.45\textwidth]{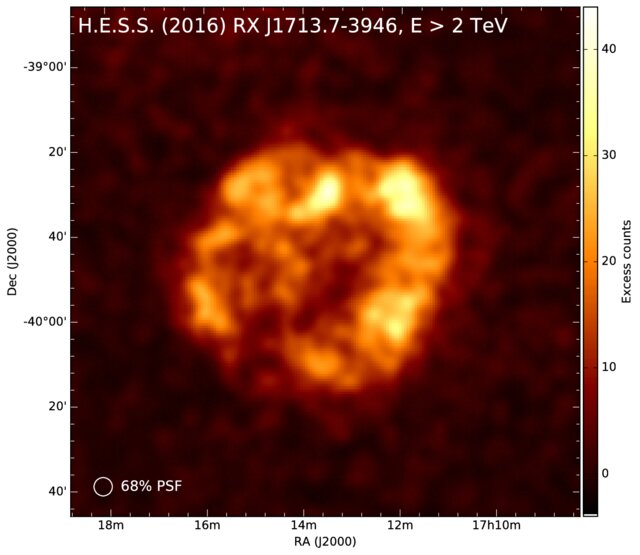}
    \includegraphics[width=0.5\textwidth]{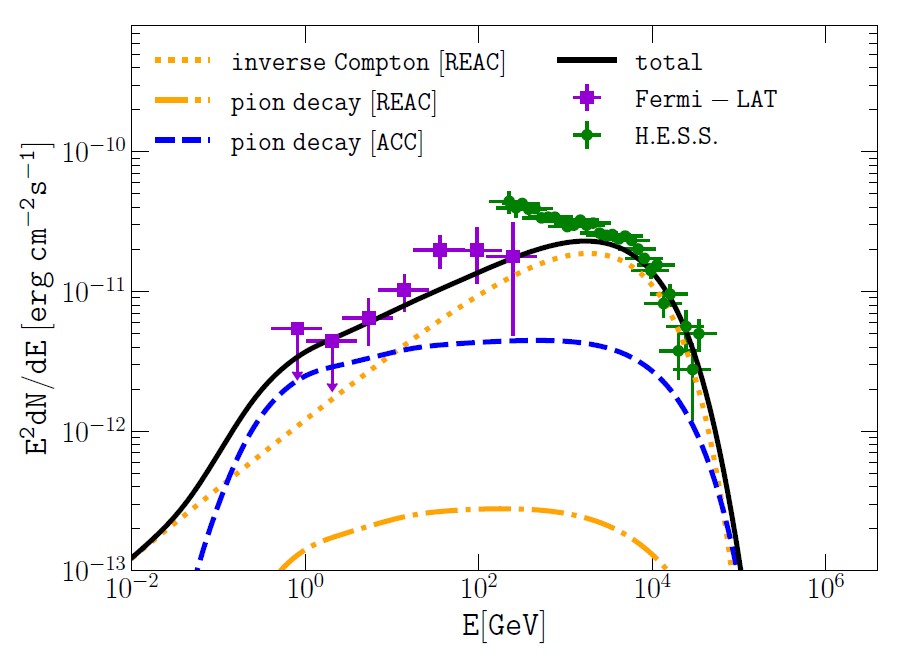}
    \caption{\textbf{Left}---\hl{The} 
 recent H.E.S.S. map of RXJ1713.7-3946, a~glowing shell of \gray\ emission coincident with the outer shock of the SNR (from \citep{Drury17}). 
    \textbf{Right}---The differential spectrum of the source obtained with H.E.S.S.  and Fermi–LAT observations. The~dotted (yellow) and dot-dashed (yellow) lines correspond to the gamma rays from re-accelerated electrons and protons. The~dashed blue line corresponds to freshly accelerated protons. The~solid black line is the sum of gamma rays from freshly accelerated protons and re-accelerated electrons (from \citep{Cristofari21_1713}).}
    \label{fig:1713}
\end{figure}

The leptonic scenario is supported by the lack of thermal X-ray emission \citep{Ellison_1713_2010,Katz_2008} and by the very good correlation between its X-ray shell \citep{Slane_1713_1999, Tanaka_1713_2008} and TeV \gray\ emission. The~very hard spectrum at GeV energy detected by Fermi-LAT \citep{Abdo11_J1713} seemed a clear indication of an Inverse Compton-dominant emission (see Figure~\ref{fig:1713}).

However, a more in-depth analysis of the environment in which the SNR is expanding stressed its non-homogeneous nature \citep{Sano_1713_2015}, particularly the presence of dense cloud cores in the northwestern part of the remnant, where there is enhanced X-ray emission. This could be a hint of the presence of an amplified magnetic field due to a shock--cloud interaction. The~clumpy nature of the RX J1713.7-3946 environment was taken into account to explain the source \gray\ spectrum with a hadronic scenario \citep{Zirakashvili_1713_2010, Gabici_1713_2014, Inoue_1713_2012, Celli_2019_tesi, Cristofari21_1713}.

For this ambiguity,  RX J1713.7-3946 is the perfect candidate for the search for neutrino emission \citep{Kappes_2007, Morlino_1713_2009, Villante_2008}. Indeed, after~the recent results of LHAASO, it is now certain that the only way we have to finally distinguish hadronic from leptonic \gray\ emission is the detection of neutrinos correlated with VHE \gray.


\subsection{IC~443}
\label{sec:IC443}

IC 443 is a middle-aged SNR (the age is thought to be about 30,000 yr~\cite{Petre88_IC443}) that belongs to the class of interacting SNRs; a system of MCs surrounds it \citep{Cornett77_IC443}. 

\begin{figure}[H]
    \includegraphics[width=0.4\textwidth]{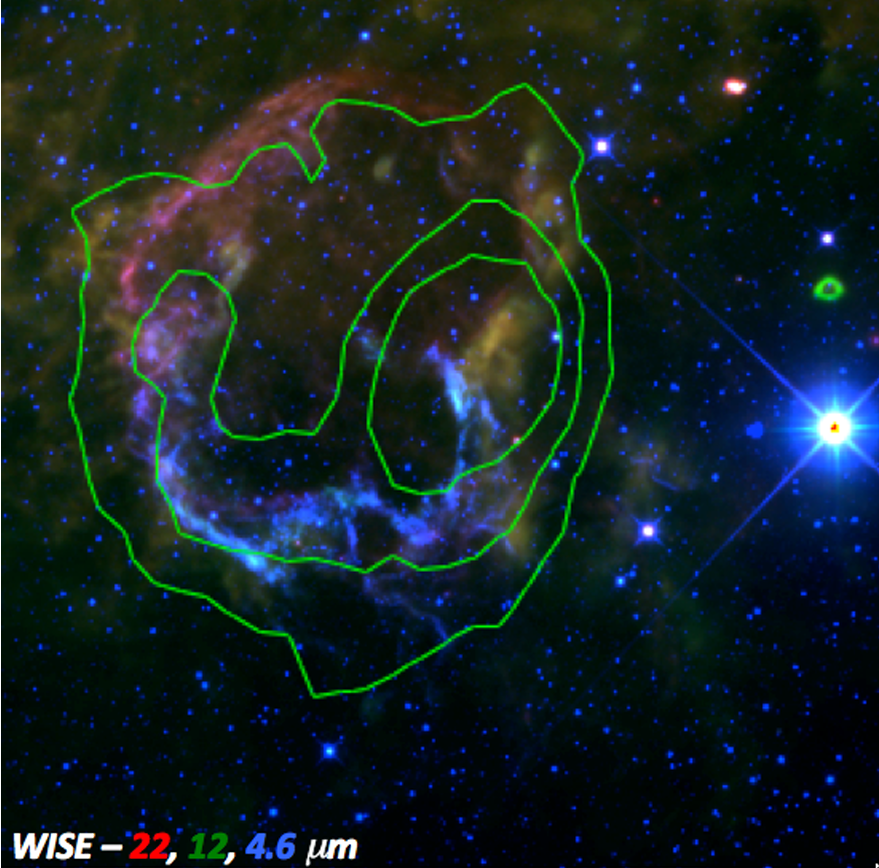}
    \includegraphics[width=0.57\textwidth]{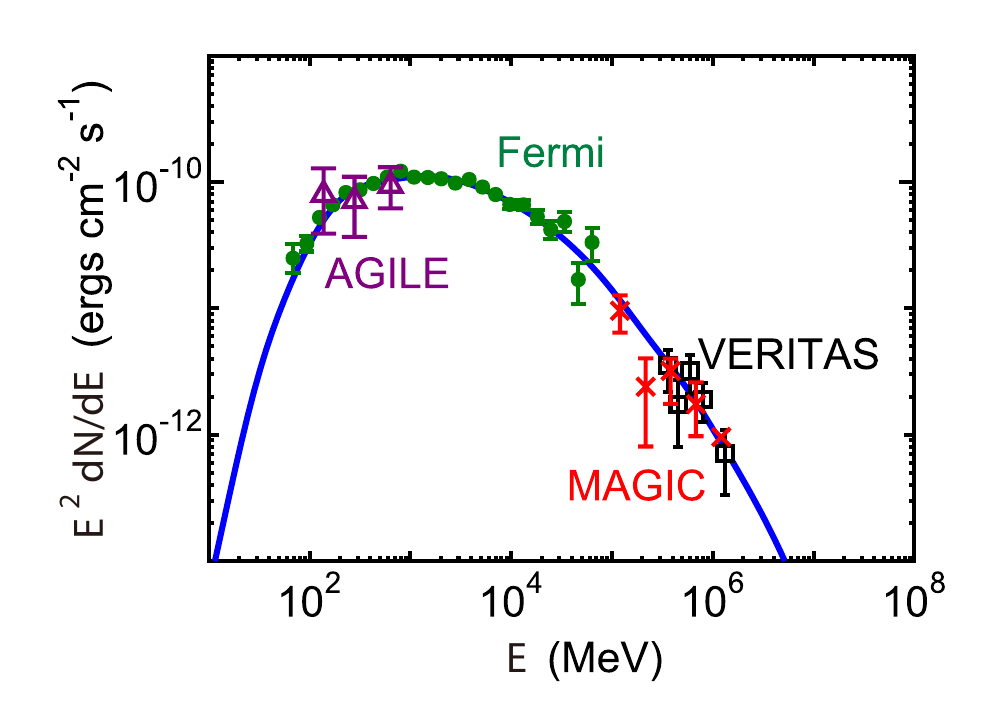}
    \caption{\hl{Morphology} 
  and spectrum  of IC 443 (from \citep{Humensky16_IC443}).} 
    \label{fig:ic443}
\end{figure}

In correspondence to the SNR-MC interaction region, \gray\ emission was detected in the TeV band by MAGIC \citep{Albert07_IC443} and VERITAS \citep{Acciari09_IC443} and in the GeV band by AGILE \citep{Tavani10_IC443} and Fermi \citep{Abdo10_IC443} (see Figure~\ref{fig:ic443}). 
The association between \gray\ emission and MCs and the~spectral features typical of pion decays indicate its hadronic origin \citep{Ackermann13}. 

\vspace{-6pt}

In the work in~\cite{Indriolo10_IC443}, an~H+3 column density near IC 443 was measured, and  a high ionization rate of \hl{2} 
 $\times$ $10^{15}$ s$^{-1}$, ﬁve times larger than the typical galactic values, was found.
A bright blob-like enhancement of the Fe I K$\alpha$ line located in the northwest and at the center of IC 443 was discovered \citep{Nobukawa18_ironLine}, likely due to low-energy CR protons accelerated in the SNR leaking into the MCs and ionizing the Fe atoms~therein.

A mass of about 1100 M$_\odot$ for the ambient gas in the [6.5, 1.5] km s$^{-1}$ range
was estimated. This estimation depends on the adopted $^{12}$CO/$^{13}$CO isotopic ratio. Still, it is established that a total molecular mass of 0.9–3.1 × 10$^3$ M$_\odot$ is available to interact with CRs via pion decay 
in a more extended~region. 


\subsection{W~44}
\label{sec:W44}

W44 is an interacting SNR lying at a distance of $\sim$3 kpc with an estimated age of $\sim$20,000 years.
Close to the remnant, there is a giant MC; the SNR-MC interaction is shown by the observation of OH maser emissions and a high $^{12}$CO(J = 2-1)-to-$^{12}$CO(J = 1-0) ratio~\citep{Yoshiike13_W44}.
In the radio wavelength, W 44 appears as a bright source that extends tens of arcmins.
The radio spectrum is a featureless power law in the range 0.02-10.7 GHz, 
with evidence of spectral variations across the shell \citep{Egreon17_SNRwithSRT, Loru19_SRT} and with no evident polarization \citep{Docara23_Megaplug}.
In X-ray, this SNR is filled with thermal emission, indicating the presence of ionized gas \citep{Okon20_W44XMM}.

In the GeV band, W 44 is one of the brightest sources in the sky. The~link between the \gray\ emission and the SNR was proposed, for~the first time, in~the work in~\cite{Esposito96_W44} on the basis of EGRET~data.

\begin{figure}[H]
    \includegraphics[width=0.8\textwidth]{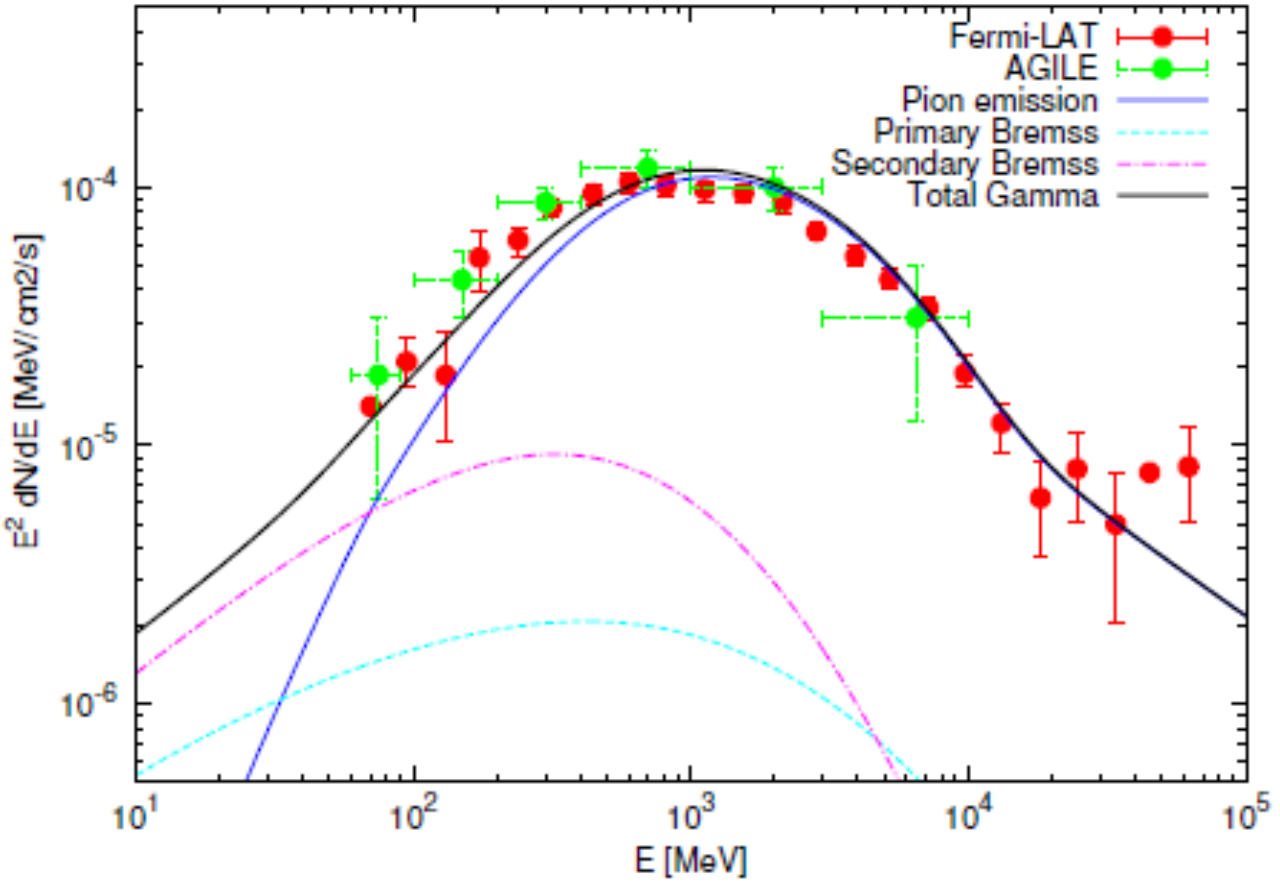}
    \caption{\hl{Gamma-ray} 
 spectrum of SNR W44 as seen by AGILE and Fermi. Hadronic and leptonic components are~indicated.}
    \label{fig:w44}
\end{figure}

It was the first SNR in whose spectrum AGILE \citep{Giuliani11_W44, Cardillo14_W44} and Fermi \citep{Ackermann13} detected a pion bump (see Figure~\ref{fig:w44}).  
Beyond a few GeV, the~spectrum becomes very soft.
Moreover, Fermi-LAT has identified two hot spots outside the SNR, aligned with it. 
These hot spots are believed to be the result of the anisotropic diffusion of CRs from the SNR, taking the form of CR clouds\citep{Gabici15_W44, Peron20_W44}.

\unskip

\subsection{Gamma~Cygni}
\label{Sec:GammaCygni}

\begin{figure}[H]
    \includegraphics[width=0.8\textwidth]{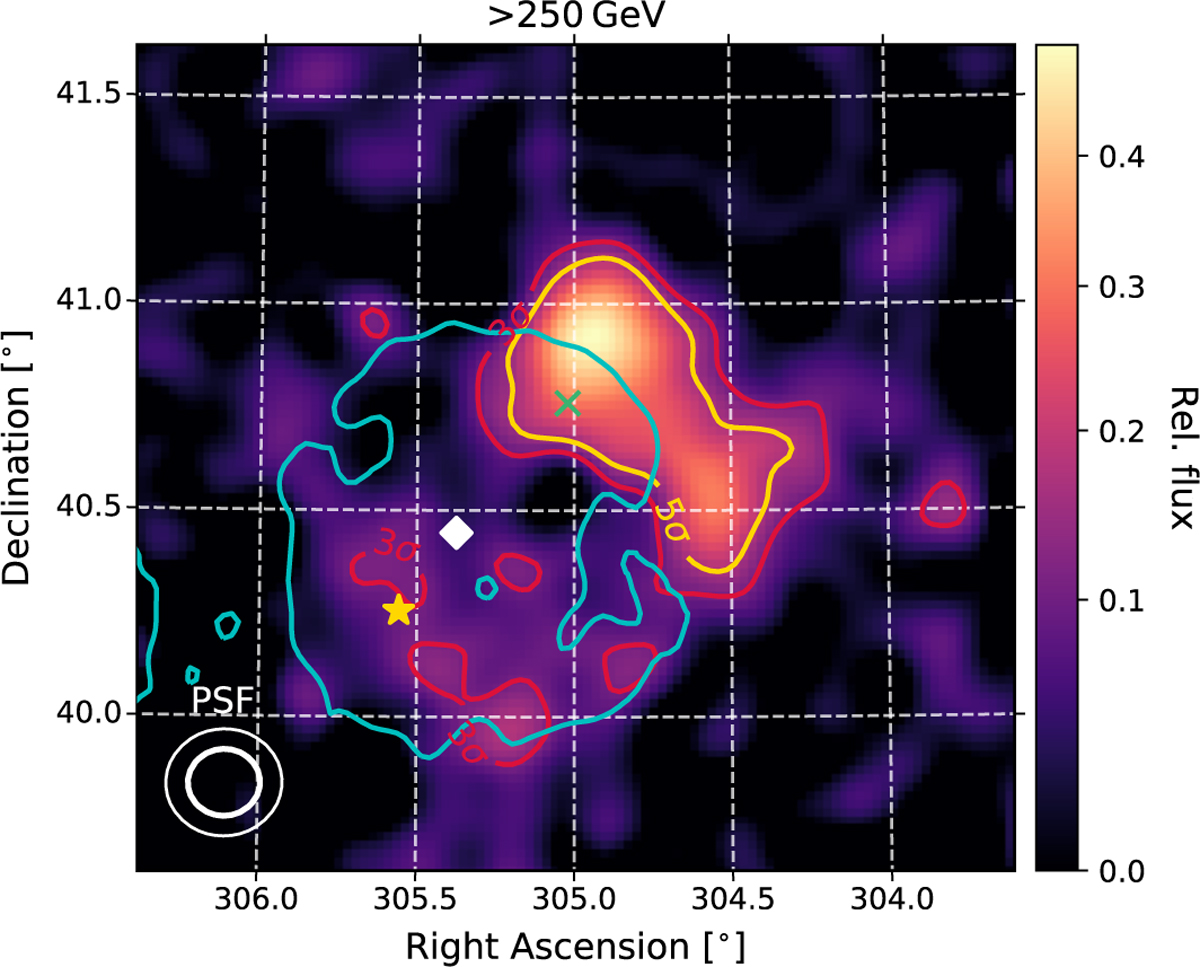}
    \caption{\hl{Gamma} 
 Cygni as seen by MAGIC \citep{magic23_gammaCygni}. The~color map indicates the gamma-ray emission, while the radio shell is shown by the blue~contour.}
    \label{fig:gammacygni}
\end{figure}

The SNR Gamma Cygni is an SNR in the constellation of Cygnus; it has an age of about 11,000~years and is characterized by
a mixed morphology (a combination of a shell-like structure and a centrally filled interior) 
and a possible interaction with MCs.
The \gray\ spectrum has been observed from a few tens of MeV to hundreds of Tev by~AGILE \citep{piano19_gammacygni}, Fermi-LAT, Veritas~\citep{abeysekara18_gammacygniVeritasFermi}, MAGIC \citep{magic23_gammaCygni} (see Figure~\ref{fig:gammacygni}), HAWC \citep{albert20_hawc3catalog}, and LHAASO \citep{Cao2024_LHAASOcat}.

The spectrum seems to classify this source as an interacting type of SNR. The~target of the accelerated protons would be given by the gas contained in the associated MCs. These are observed in the radio band with a characteristic speed of 40--50 km s$^{-1}$ and collect a mass of approximately $10^4$ $M_{\odot}$.
The recent LHAASO observation of emission up to energies of about 300 TeV makes this object interesting for understanding the ability of an SNR to accelerate particles up to~VHEs.

\unskip

\subsection{W~28}
\label{Sec:W28}

W 28 is an SNR that provides a unique opportunity to investigate the diffusion of CRs in an astrophysical environment. 
It is approximately 30,000 years old and lies at an estimated distance between 2 and 3.5 kpc.
In continuous radio, it has a crescent shape, and the non-thermal spectrum indicates the presence of relativistic electrons in the shell \citep{dubner00}.
W 28 provides a unique opportunity to investigate the diffusion of CRs in an astrophysical environment. Indeed, two giant MCs with masses of a few $10^4$ $M_{\odot}$ appear near the SNR at compatible distances~\citep{mizuno04}.
One cloud partially overlaps with the SNR shell, while the other is located approximately 0.5 degrees away from the shell.
The system can be modeled as having one cloud in direct contact with the SNR shock and another cloud about 10--15 parsecs away, assuming that the three objects are at the same distance and interacting with each other.
Evidence of an SNR-MC system interaction comes from observations of maser emission and~high ionization levels \citep{frail94}.

W 28 was one of the first SNRs to be indicated as a \gray\ source in the Cos B data~\citep{pollock85_w28CosB}.
H.E.S.S. \citep{aharonian08_W28}, AGILE \citep{giuliani10_w28}, and Fermi-LAT \citep{abdo10_w28Fermi} observations of this object measured a \gray\ spectrum that is typical of an interacting SNR, namely, a peak below 1 GeV and a very soft spectrum above it.
It is interesting to note that the peak of the \gray\ spectrum in the cloud in direct contact with the shell is at different energies than the one estimated from the other clouds. This suggests that diffusion can have an important role in this~system.


\section{Conclusions}

For the past fifty years, SNRs have been one of the main targets of \gray\ missions. 
Today, about thirty SNRs have been observed with certainty in \grays\ by~satellites in the GeV band, by~Cherenkov telescopes in the TeV band, and~more recently, by hybrid detectors like LHAASO and HAWC in the PeV band. 
Their study has provided answers but also raised questions.
For instance, their ability to accelerate CRs (electrons and hadrons) has been demonstrated. At~the same time, their spectra and morphologies have proved to be non-trivial to interpret.  
Another open question concerns the maximum energy at which SNRs can accelerate particles. 
Some of these objects have an observed spectrum that extends 
beyond 100 TeV, confirming that these objects may be capable of accelerating particles up to about 1 PeV. 
However, they represent a small minority of both the LHAASO sources and the SNRs observed in the \gray\  regime.
Overall, after~almost half a century of study, it is still unclear whether or not their contribution to galactic CR acceleration is dominant compared to other sources.
To answer these questions, a~new generation of \gray\ instruments will come into operation in the next few years, in~particular, instruments such as ASTRI Mini-Array and CTA, which will make it possible to better interpret the VHE/UHE observations that LHAASO, HAWC, and then the future SWGO will make of these objects and will finally allow us to know which astrophysical system supplies the Galaxy with its CR~population.


\vspace{6pt} 




\authorcontributions{Conceptualization, A.G.; resources, M.C. and A.G.; writing---original draft preparation, A.G.; writing---review and editing, M.C. and A.G. All authors have read and agreed to the published version of the~manuscript.}

\funding{This research received no external~funding}

\acknowledgments{\hl{We} 
 are grateful to the reviewers for their useful comments, which have enhanced the quality of our review.
This research has made use of the TeVCat online source \hl{catalog}   
 (\url{http://tevcat.uchicago.edu}) 
and the SNRcat online catalog (\url{http://snrcat.physics.umanitoba.ca}), both accessed on 1 January 2024.
}

\conflictsofinterest{The authors declare no conflicts of~interest.} 

\abbreviations{Abbreviations}{
The following abbreviations are used in this manuscript:

\noindent 
\begin{tabular}{@{}ll}
ALPs & Axion-like particles\\
AS-$\gamma$ & Air shower \gray{} array\\
ASTRI & Astrofisica con Specchi a Tecnologia Replicante Italiana\\
C.L. & Confidence Limit\\
CTAO & \v{C}erenkov Telescope Array Observatory\\
DM & Dark matter\\
EAS  & Extended air shower arrays\\
ESA & European Space Agency\\
EBL & Extra-galactic background light\\
E-HBL & Extreme high-peaked BL Lacs\\
eROSITA & Extended Roentgen survey with an imaging telescope array\\
FOV & Field of view\\
FR & Fanaroff–Riley galaxies \\ 
FSRQ & Flat-spectrum radio quasar\\
GASP & GLAST-AGILE Support Programme\\
GRB & Gamma-ray burst\\
GW & Gravitational wave\\
HAWC & High-Altitude Water \v{C}erenkov Observatory\\
HB & Hadron Beam\\
HBL & High-peaked BL Lacs \\ 
HE & High energy\\
H.E.S.S. & High-Energy Stereoscopic System\\
IAC & Instituto de Astrof\'isica de Canarias\\
IACT & Imaging Atmospheric \v{C}erenkov Telescope arrays\\
IBL & Intermediate-peaked BL Lacs \\  
IGMF & Inter-galactic magnetic field\\
IR & Infra-red\\
IXPE & Imaging X-ray Polarimetry Explorer\\
LHAASO & Large High-Altitude Air Shower Observatory\\
LIV & Lorentz invariance violation\\\end{tabular}
}

\abbreviations{}{

\noindent 
\begin{tabular}{@{}ll}
LST & Large-sized telescope\\
MAGIC & Major Atmospheric Gamma-Ray Imaging \v{C}erenkov telescopes\\
MST & Medium-sized telescope\\
NASA & National Aeronautics and Space Administration\\
SBG &  Star-bursting galaxies \\
SC & Schwarzschild--Couder\\
SII & Stellar intensity interferometry\\
SRT & Sardinia radio telescope\\
SST & Small-sized telescope\\
TNG & Telescopio Nazionale Galileo\\
VERITAS & Very Energetic Radiation Imaging Telescope Array System\\
VHE & Very high energy\\
WEBT & Whole-Earth Blazar Telescope\\
\end{tabular}
}

\begin{adjustwidth}{-\extralength}{0cm}

\reftitle{References}

\PublishersNote{}
\end{adjustwidth}
\end{document}